\documentclass[aps,float,prd,psfig]{revtex4}
\input epsf
\usepackage{graphicx}% Include figure files
\newcommand{\beq}{\begin{equation}}
\newcommand{\beqa}{\begin{eqnarray}}
		  \newcommand{\eeq}{\end{equation}}
\newcommand{\eeqa}{\end{eqnarray}}

\newcommand{\vect}[1]{\mbox{\boldmath${#1}$}}
\newcommand{\lmk}{\left(}
\newcommand{\rmk}{\right)}
\newcommand{\lnk}{\left\{ }
\newcommand{\rnk}{\right\} }
\newcommand{\lkk}{\left[}
\newcommand{\rkk}{\right]}
\newcommand{\lla}{\left\langle}

\newcommand{\rra}{\right\rangle}

\newcommand{\vex}{{\vect x}}

\newcommand{\ven}{\vect n}
\newcommand{\vem}{\vect m}

\newcommand{\veu}{{\vect u}}
\newcommand{\vev}{{\vect v}}

\newcommand{\ved}{{\vect d}}
\newcommand{\ve}{{\vect e}}

\begin{document}
%\baselineskip 8mm
%%%%%%%%%%%%%%%%%%%%%%%%%%%%%%%%%%%%%%%%%%%%%%%%%%%%%%%%%%%%%%%%%%%%%%%%%%
%%%%%%%%%%%%%%%%%%%%%%%%%%%%%%%%%%%%%%%%%%%%%%%%%%%%%%%%%%%%%%%%%%%%%%%%%%
\title{
Measuring a Parity Violation Signature in the Early Universe via 
   Ground-based Laser Interferometers } 
%%%%%%%%%%%%%%%%%%%%%%%%%%%%%%%%%%%%%%%%%%%%%%%%%%%%%%%%%%%%%%%%%%%%%%%%%%
%%%%%%%%%%%%%%%%%%%%%%%%%%%%%%%%%%%%%%%%%%%%%%%%%%%%%%%%%%%%%%%%%%%%%%%%%%
%
%
%
%%%%%%%%%%%%%%%%%%%%%%%%%%%%%%%%%%%%%%%%%%%%%%%%%%%%%%%%%%%%%%%%%%%%%%%%%%
\author{Naoki Seto$^{1,2}$ and Atsushi Taruya$^3$}
%%%%%%%%%%%%%%%%%%%%%%%%%%%%%%%%%%%%%%%%%%%%%%%%%%%%%%%%%%%%%%%%%%%%%%%%%%
\affiliation{$^1$National Astronomical Observatory, 2-21-1
Osawa, Mitaka, Tokyo, 181-8588, Japan\\ 
$^2$Department of Physics and Astronomy, 4186 Frederick Reines
Hall, University of California, Irvine, CA 92697\\
$^3$Research Center for the Early Universe, School of Science, The University of Tokyo, Tokyo 113-0033, Japan
}
%%%%%%%%%%%%%%%%%%%%%%%%%%%%%%%%%%%%%%%%%%%%%%%%%%%%%%%%%%%%%%%%%%%%%%%%%%
\date{\today}
%
%
%
%
%
%
%%%%%%%%%%%%%%%%%%%%%%%%%%%%%%%%%%%%%%%%%%%%%%%%%%%%%%%%%%%%%%%%%%%%%%%%%%
\begin{abstract}
 We show that pairs of widely separated  interferometers 
 are advantageous for measuring the Stokes parameter $V$  of a stochastic
 background of gravitational waves. This parameter characterizes asymmetry of 
 amplitudes of right- and left-handed  waves and generation of the asymmetry 
 is closely related to parity violation in the early universe. 
 The advantageous pairs include  LIGO (Livingston)-LCGT and AIGO-Virgo
 that are relatively insensitive to $\Omega_{\rm\scriptscriptstyle GW}$ 
(the simple intensity of 
 the background). Using at least three detectors,  information of the 
 intensity $\Omega_{\rm\scriptscriptstyle GW}$ and the degree of asymmetry 
 $V$ can be separately measured. 
\end{abstract}
%%%%%%%%%%%%%%%%%%%%%%%%%%%%%%%%%%%%%%%%%%%%%%%%%%%%%%%%%%%%%%%%%%%%%%%%%%

\maketitle

%%%%%%%%%%%%%%%%%%%%%%%%%%%%%%%%%%%%%%%%%%%%%%%%%%%%%%%%%%%%%%%%%%%%%%%%%%
%%%%%%%%%%%%%%%%%%%%%%%%%%%%%%%%%%%%%%%%%%%%%%%%%%%%%%%%%%%%%%%%%%%%%%%%%%
\section{Introduction}
%\underline{\em 1)  Introduction}
%%%%%%%%%%%%%%%%%%%%%%%%%%%%%%%%%%%%%%%%%%%%%%%%%%%%%%%%%%%%%%%%%%%%%%%%%%
%%%%%%%%%%%%%%%%%%%%%%%%%%%%%%%%%%%%%%%%%%%%%%%%%%%%%%%%%%%%%%%%%%%%%%%%%%

Stochastic background of gravitational waves is one of the most important
targets for gravitational wave astronomy.  
In the last decade, the detection threshold for the background has been 
rapidly improved around $\sim 100$Hz by 
continuous upgrades of ground-based  interferometers \cite{Abbott:2005ez}. 
This trend will be continued 
 with advent of next-generation interferometers currently planned
worldwide, such as advanced LIGO \cite{adv} and LCGT
\cite{Kuroda:1999vi}. Due to the weakness of 
gravitational interaction, our universe is transparent to the
background up to very early epoch, and we might uncover interesting 
nature of the universe at extremely high-energy scales, 
through observational studies of the stochastic background. 
To extract the information as much as possible, we need to 
characterize the background efficiently in a model independent manner,
and  investigation 
beyond simple spectral analysis might yield a great discovery. 
In this respect, circular polarization degree,  
which describes the asymmetry between the amplitudes of right- and 
left-handed waves, may be a fundamental characteristic of the 
background to probe the early universe. 
Because the parity transformation relates these two polarization modes, 
the asymmetry in the stochastic gravitational waves 
directly reflects a parity violation in the early universe,  for instance,
generated through the gravitational  Chern-Simons term 
 (e.g., \cite{Alexander:2004us}). 
In other words, one can detect a signature 
of parity violation by measuring the circular polarization degree of a 
gravitational wave background. 
Since the observed universe is highly isotropic and homogeneous, we shall 
focus on the monopole component of the circular polarization as our primary 
target, and report principle aspects for its measurement with a 
network of ground-based interferometers (see  \cite{Lue:1998mq,Caprini:2003vc} for CMB polarization
and \cite{Seto:2006hf} for space missions).

%%%%%%%%%%%%%%%%%%%%%%%%%%%%%%%%%%%%%%%%%%%%%%%%%%%%%%%%%%%%%%%%%%%%%%%%%%
%%%%%%%%%%%%%%%%%%%%%%%%%%%%%%%%%%%%%%%%%%%%%%%%%%%%%%%%%%%%%%%%%%%%%%%%%%
\section{Circular polarization}
%\underline{\em 2)  Circular polarization}
%%%%%%%%%%%%%%%%%%%%%%%%%%%%%%%%%%%%%%%%%%%%%%%%%%%%%%%%%%%%%%%%%%%%%%%%%%
%%%%%%%%%%%%%%%%%%%%%%%%%%%%%%%%%%%%%%%%%%%%%%%%%%%%%%%%%%%%%%%%%%%%%%%%%%

Let us first describe circular polarization of a gravitational 
wave background. We use a plane wave expansion of the background as
\cite{Flanagan:1993ix,Allen:1997ad} 
%%%%%%%%%%%%%%%%%%%%%%%%%%%%%%%%%%%%%%%%%%%%%%%%%%%%%%%%%%%%%%%%%%%%%%%%%%
\beq
h_{ij}(t,\vex)=\sum_{P=+,\times} \int^{\infty}_{-\infty} df \int_{S^2} d\ven~
h_P(f,\ven) e^{2\pi i f (-t+\ven \cdot \vex) } \ve^P_{ij}(\ven).
\label{plane}
\eeq
%%%%%%%%%%%%%%%%%%%%%%%%%%%%%%%%%%%%%%%%%%%%%%%%%%%%%%%%%%%%%%%%%%%%%%%%%%
Here, the amplitude $h_P$ is the mode coefficient that is stochastic and 
random variable. 
The bases for transverse-traceless tensor $\ve^P$ $(P=+,\times)$ 
are given as 
$\ve^+_{}={\hat \ve}_\theta \otimes {\hat \ve}_\theta- {\hat \ve}_\phi
\otimes  {\hat \ve}_\phi$ and $\ve^\times_{}={\hat \ve}_\theta \otimes
{\hat 
\ve}_\phi+{\hat  
\ve}_\phi \otimes {\hat \ve}_\theta$ with unit vectors $ {\hat
\ve}_\theta$ and $ {\hat \ve}_\phi$. These vectors are 
normal to the propagation
direction $\ven$, associated with  a right-handed Cartesian 
coordinate as usual. As an alternative characterization, 
we can use the circular polarization bases 
$\ve^R=(\ve^++i\ve^\times)/\sqrt2$ (right-handed mode) and 
 $\ve^L=(\ve^+-i\ve^\times)/\sqrt2$ (left-handed mode) for the plane wave
expansion (\ref{plane}). The corresponding amplitudes $h_{R}$ and $h_{L}$ 
are given by 
$h_R=(h_+-i\,h_\times)/\sqrt2$  and $h_L=(h_++i\,h_\times)/\sqrt2$. 
The ensemble average of their amplitudes is classified as 
%%%%%%%%%%%%%%%%%%%%%%%%%%%%%%%%%%%%%%%%%%%%%%%%%%%%%%%%%%%%%%%%%%%%%%%%%%
\beq
\left( \begin{array}{@{\,}c@{\,}}
             \lla h_R(f,\ven) h_R(f',\ven')^* \rra  \\
              \lla h_L(f,\ven) h_L(f',\ven')^* \rra \\ 
           \end{array} \right)
=\frac{\delta_{\ven,\ven'}\delta_{f,f'}}{4\pi}\left( 
           \begin{array}{@{\,}c@{\,}}
           I(f,\ven)+V(f,\ven)  \\
           I(f,\ven)-V(f,\ven)  \\ 
           \end{array} \right)  \label{matrix}
\eeq
%%%%%%%%%%%%%%%%%%%%%%%%%%%%%%%%%%%%%%%%%%%%%%%%%%%%%%%%%%%%%%%%%%%%%%%%%%
with  the functions $\delta_{Y,Z}$ being delta functions. 
%The bracket $\lla \cdots \rra$ stands for an ensemble average. 
In the above expression, the real function $V$ characterizes the 
asymmetry between the amplitudes of right- and  the left-handed waves, while 
the function $I(\ge |V|)$ represents their total amplitude. 
Note that the other combinations such as $\lla h_R h_L^* \rra$ and 
$\lla h_L h_R^* \rra$ describe the 
linear polarization mode and are proportional to $Q\pm i\,U$, 
which constitute the well-known Stokes parameter, 
together with the $I$- and $V$- modes  
(see e.g., \cite{radipro} for electromagnetic counterpart). 
In this paper, we do not study the linear polarization $Q\pm i\,U$, since 
they do not have an isotropic component. 
We will focus on the detectability of the isotropic 
components $I(f)$ and $V(f)$ as our primary target. 
Using the normalized logarithmic energy density of the background 
$\Omega_{\rm \scriptscriptstyle GW}(f)$ \cite{Flanagan:1993ix,Allen:1997ad}, 
the two functions $I$ and $V$ are expressed as  
%%%%%%%%%%%%%%%%%%%%%%%%%%%%%%%%%%%%%%%%%%%%%%%%%%%%%%%%%%%%%%%%%%%%%%%%%%
\beq
I(f)=\frac{\rho_{\rm c}}{4\pi f^3}\,\,\Omega_{\rm\scriptscriptstyle GW}(f),
\quad
V(f)=\frac{\rho_{\rm c}}{4\pi f^3}\,\,\Omega_{\rm\scriptscriptstyle GW}(f)\,
\Pi(f),
\eeq
%%%%%%%%%%%%%%%%%%%%%%%%%%%%%%%%%%%%%%%%%%%%%%%%%%%%%%%%%%%%%%%%%%%%%%%%%%
where $\rho_{\rm c}$ is the critical density of the Universe,  
$\rho_{\rm c}=3H_0^2/8\pi$ with $H_0=70h_{70}$\,km/sec/Mpc 
being the Hubble parameter. 
The ratio $\Pi(f)=V(f)/I(f)$ characterizes the circular polarization degree. 
For simplicity, we assume the flat spectra, 
$\Omega_{\rm \scriptscriptstyle GW}(f)\propto f^0$ 
and $\Pi(f)\propto f^0$ as our fiducial model. Thus, our main interest is the simultaneous 
determinations or constraints on the parameters 
$\Omega_{\rm\scriptscriptstyle GW}$ and $\Pi$.

We next consider how to detect the isotropic components of $I$- and 
$V$-modes with laser interferometers. Let us recall that the 
output signal $s_a$ of a detector $a$ at the position $\vex_a$ is written as 
$
s_a(f) =\sum_{P=+,\times}\int_{S^2} d\ven \,h_P(f,\ven)\, F^P_{a}(\ven, f)
\,e^{i\,\,2\pi \, f\, \ven\cdot\vex_a}. 
$
Here, the function  $F_a^P$ is the beam pattern function and it 
represents the response of the detector to a polarization mode $\ve^{P}$. 
Provided the data streams $s_a$ and $s_b$ taken from two detectors 
$a$ and $b$, the detection of stochastic signals can be achieved by 
taking a cross-correlation,   
$
 \lla  s_a(f) s_b(f')^* \rra \equiv C_{ab}(f) \delta_{f,f'}.
$
Keeping the signals from the isotropic components,  
the correlation signal $C_{ab}(f)$ is written as 
%%%%%%%%%%%%%%%%%%%%%%%%%%%%%%%%%%%%%%%%%%%%%%%%%%%%%%%%%%%%%%%%%%%%%%%%%%
\beq
 C_{ab}(f)=
\gamma_{I,ab}(f)I(f)+\gamma_{V,ab}(f)V(f),
\eeq 
%%%%%%%%%%%%%%%%%%%%%%%%%%%%%%%%%%%%%%%%%%%%%%%%%%%%%%%%%%%%%%%%%%%%%%%%%%
where the quantity $\gamma_I$ is the overlap  function
given by \cite{Flanagan:1993ix,Allen:1997ad} 
%%%%%%%%%%%%%%%%%%%%%%%%%%%%%%%%%%%%%%%%%%%%%%%%%%%%%%%%%%%%%%%%%%%%%%%%%%
\beq
\gamma_{I,ab}(f)=\frac{5}{8\pi}\int_{S^2} d\ven  \lkk  \lnk
F_a^+F_{b}^{+*}+
F_a^\times F_{b}^{\times*} \rnk e^{i\,y\,\ven\vem} \rkk, \label{gi1}
\eeq
%%%%%%%%%%%%%%%%%%%%%%%%%%%%%%%%%%%%%%%%%%%%%%%%%%%%%%%%%%%%%%%%%%%%%%%%%%
with $y\equiv2\pi f\,D/c$. Here, we have expressed 
$\vex_a-\vex_b$ as $D \vem$ ($D$:\,\,distance, $\vem$:\,\,unit vector). 
Similarly, the function $\gamma_{V,ab}(f)$ is 
obtained by replacing the kernel $[\cdots]$ in Eq.~(\ref{gi1}) 
with $\lkk i \lnk 
F_a^+F_{b}^{\times*}-
F_a^\times F_{b}^{+*} \rnk e^{iy \ven\vem}  \rkk $.

%%%%%%%%%%%%%%%%%%%%%%%%%%%%%%%%%%%%%%%%%%%%%%%%%%%%%%%%%%%%%%%%%%%%%%%%%%
%%%%%%%%%%%%%%%%%%%%%%%%%%%%%%%%%%%%%%%%%%%%%%%%%%%%%%%%%%%%%%%%%%%%%%%%%%
\section{Overlap functions for ground based detectors}
%\underline{\em 3)  Overlap functions for ground based detectors}
%%%%%%%%%%%%%%%%%%%%%%%%%%%%%%%%%%%%%%%%%%%%%%%%%%%%%%%%%%%%%%%%%%%%%%%%%%
%%%%%%%%%%%%%%%%%%%%%%%%%%%%%%%%%%%%%%%%%%%%%%%%%%%%%%%%%%%%%%%%%%%%%%%%%%

Now, specifically consider the response of an L-shaped 
interferometer $a$ on the 
Earth. We assume that the detector has two orthogonal arms with equal 
arm-length. 
\if0%%%%%%%%%%%%%%%%
In reality an interferometer might has an opening angle different from
$90^\circ$. For example the  angle of GEO600 is $94.3^\circ$. But
its response can be effectively regarded  as an interferometer with
$90^\circ$ opening angle.
\fi%%%%%%%%%%%%%%%%
Denoting the unit vectors parallel to the two arms by $\veu$ and 
$\vev$, the beam pattern function takes a simple form as 
$F_a^P=\ved_a:\ve^P(\ven)$ with 
$\ved_a=({\veu} \otimes {\veu}- {\vev} \otimes  {\vev})/2$, 
where the colon represents a double contraction.  
This expression is always valid as long as the wavelength of the 
gravitational waves for our interest is much longer than 
the arm-length of the detectors.  
\if0%%%%%%%%%%%%%%%%%% 
Table \ref{tab:detectors} provides a list of the positions and the 
orientations for ongoing (and planned) kilometer-size interferometers 
(see e.g., \cite{det}).  We use a spherical coordinate system
$(\theta,\phi)$ in which the north pole is located at 
$\theta=0^{\circ}$, with $\phi$ representing the longitude. 
The angle $\alpha$  characterizes the orientation of the detector between 
the local east direction and the bisecting line of the two arms 
measured counterclockwise.
\fi%%%%%%%%%%%%%%%%%%%%%%%
In this paper we study the following five ongoing (and planned)
kilometer-size interferometers as concrete examples; AIGO(A), LCGT(C),
LIGO-Hanford(H), LIGO-Livingston(L) and Virgo(V) (see {\it e.g.}
\cite{det} for their basic information). 
 Hereafter, 
we mainly use their abbreviations (A,C,H,L,V).

\if0
%%%%%%%%%%%%%%%%%%%%%%%%%%%%%%%%%%%%%%%%%%%%%%%%%%%%%%%%%%%%%%%%%%%%%%%%%%
%%%%%%%%%%%%%%%%%%%%%%%%%%%%%%%%% Table 1 %%%%%%%%%%%%%%%%%%%%%%%%%%%%%%%%
%%%%%%%%%%%%%%%%%%%%%%%%%%%%%%%%%%%%%%%%%%%%%%%%%%%%%%%%%%%%%%%%%%%%%%%%%%
\begin{table}[!tb]
%\begin{ruledtabular}
\begin{tabular}{lccc}
detector & $\theta$ & $\phi$ & $\alpha$ \\
\hline
\ AIGO (A) & 121.4 & 115.7 & -45.0\\
\ LCGT (C) & 53.6 & 137.3 & 70.0 \\
\ LIGO\ Hanford (H) & 43.5 & -119.4 & 171.8 \\
\ LIGO\ Livingston (L) & 59.4 & -90.8 & 243.0 \\
\ Virgo (V) & 46.4 & 10.5 & 116.5 
\end{tabular}
%\end{ruledtabular}
\caption{Positions $(\theta,\phi)$ and orientation angles $\alpha$ of
 detectors (in units of degree).   
\label{tab:detectors}}
\end{table}
%%%%%%%%%%%%%%%%%%%%%%%%%%%%%%%%%%%%%%%%%%%%%%%%%%%%%%%%%%%%%%%%%%%%%%%%%%
%%%%%%%%%%%%%%%%%%%%%%%%%%%%%%%%%%%%%%%%%%%%%%%%%%%%%%%%%%%%%%%%%%%%%%%%%%
\fi

\begin{figure}
  \begin{center}
\epsfxsize=4.cm
\begin{minipage}{\epsfxsize} \epsffile{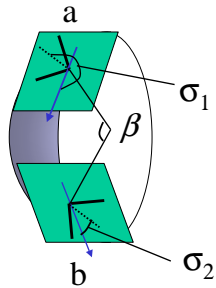} \end{minipage}
 \end{center}
  \caption{  Geometrical configuration of ground-based detectors $a$ and 
    $b$ for the cross-correlation analysis. Detector planes are 
    tangential to the Earth. Two detectors 
 $a$ and $b$ are separated by the angle $\beta$ measured from the center
 of the Earth. The angles 
 $\sigma_1$ and $\sigma_2$ describe the orientation of bisectors of
 interferometers in a counter-clockwise manner relative to the great 
 circle joining two sites. \label{fig:detector_config}
 }
%\label{f2}
\end{figure}

%We can explicitly calculate correlation $C_{ab}$ of data streams for two
%detectors $a$ and $b$. 
For the isotropic component of the stochastic background, only the 
relative configuration of two detectors is relevant with the correlation 
signal $C_{ab}$ and we do not care about the overall rotation. 
Hence, the sensitivity of each pair of detectors to the stochastic background 
 can be characterized by the three angular parameters 
$(\beta,\sigma_1,\sigma_2)$ shown in Fig.~\ref{fig:detector_config}. Here, 
$\beta$ is the separation angle between two detectors measured 
from the center of the Earth. The angle $\sigma_1$ ($\sigma_2$) is the
orientation of the bisector of two arms for detector $a$ ($b$) 
measured in counter-clockwise manner relative to the great 
circle connecting $a$ and $b$. Their distance is given by
$
D=2R_E\sin(\beta/2)
$
($R_E=6400$km : the radius of the Earth), which 
determines a characteristic frequency $f_D\equiv c/(2\pi\,D)$ for the overlap
functions. 
%We have $f_D=91$Hz, 33Hz and 23Hz
%for $\beta=30^\circ$, $90^\circ$ and $180^\circ$ respectively.
Following Ref.\cite{Flanagan:1993ix}, we define the  angles 
%%%%%%%%%%%%%%%%%%%%%%%%%%%%%%%%%%%%%%%%%%%%%%%%%%%%%%%%%%%%%%%%%%%%%%%%%%
\beq
\Delta\equiv ({\sigma_1+\sigma_2})/2,~~~\delta\equiv ({\sigma_1-\sigma_2})/2.
\eeq
%%%%%%%%%%%%%%%%%%%%%%%%%%%%%%%%%%%%%%%%%%%%%%%%%%%%%%%%%%%%%%%%%%%%%%%%%%
The geometrical information about pairs of detectors 
among the five interferometers is presented in Table \ref{tab:angles}.

%%%%%%%%%%%%%%%%%%%%%%%%%%%%%%%%%%%%%%%%%%%%%%%%%%%%%%%%%%%%%%%%%%%%%%%%%%
%%%%%%%%%%%%%%%%%%%%%%%%%%%%%%%%% Figure 1 %%%%%%%%%%%%%%%%%%%%%%%%%%%%%%%
%%%%%%%%%%%%%%%%%%%%%%%%%%%%%%%%%%%%%%%%%%%%%%%%%%%%%%%%%%%%%%%%%%%%%%%%%%
%%%%%%%%%%%%%%%%%%%%%%%%%%%%%%%%%%%%%%%%%%%%%%%%%%%%%%%%%%%%%%%%%%%%%%%%%%
%%%%%%%%%%%%%%%%%%%%%%%%%%%%%%%%%%%%%%%%%%%%%%%%%%%%%%%%%%%%%%%%%%%%%%%%%%

%%%%%%%%%%%%%%%%%%%%%%%%%%%%%%%%%%%%%%%%%%%%%%%%%%%%%%%%%%%%%%%%%%%%%%%%%%
%%%%%%%%%%%%%%%%%%%%%%%%%%%%%%%%% Table 2 %%%%%%%%%%%%%%%%%%%%%%%%%%%%%%%%
%%%%%%%%%%%%%%%%%%%%%%%%%%%%%%%%%%%%%%%%%%%%%%%%%%%%%%%%%%%%%%%%%%%%%%%%%%
\begin{table}[!t]
%\begin{ruledtabular}
\begin{tabular}{l|c|c|c|c|c}
  & A &  C & H & L & V \\
\hline
\ A &* & 70.8$^{\circ}$,~-0.61 & 135.6$^{\circ}$,~-0.82 &157.3$^{\circ}$,~-0.88  
&121.4$^{\circ}$,~0.23 \\
\hline
\ C & -0.58,~0.81 &*  & 72.4$^{\circ}$,~1.00  &99.2$^{\circ}$,~-0.98 & 
86.6$^{\circ}$,~-0.43 \\
\hline
\ H &-1.00,~-0.007  &-0.21,~0.98  & * &  27.2$^{\circ}$,~-1.00 & 
79.6$^{\circ}$,~-0.43 \\
\hline
\ L &0.99,~0.15  &0.04,~-1.00  & -0.36,~-0.93   & * &76.8$^{\circ}$,~-0.29 \\
\hline
\ V &-0.45,~-0.89  & 0.92,~0.38 & -0.76,~-0.65  &0.89,~-0.46 &  *
\end{tabular}
%\end{ruledtabular}
\caption{Upper right $(\beta, \cos(4\delta))$. Lower left 
$(\cos (4\Delta),~\sin(4\Delta))$. \label{tab:angles}}
\end{table}
%%%%%%%%%%%%%%%%%%%%%%%%%%%%%%%%%%%%%%%%%%%%%%%%%%%%%%%%%%%%%%%%%%%%%%%%%%
%%%%%%%%%%%%%%%%%%%%%%%%%%%%%%%%%%%%%%%%%%%%%%%%%%%%%%%%%%%%%%%%%%%%%%%%%%

In the expression (\ref{gi1}), the angular integral can be performed 
analytically with explicit forms of the pattern functions. 
A long but straightforward calculation leads to \cite{Flanagan:1993ix}  
%%%%%%%%%%%%%%%%%%%%%%%%%%%%%%%%%%%%%%%%%%%%%%%%%%%%%%%%%%%%%%%%%%%%%%%%%%
\beq
\gamma_{I,ab}= \Theta_1(y,\beta)\,\cos(4\delta)+
\Theta_2(y,\beta) \,\cos(4\Delta), 
\label{gi}
\eeq
%%%%%%%%%%%%%%%%%%%%%%%%%%%%%%%%%%%%%%%%%%%%%%%%%%%%%%%%%%%%%%%%%%%%%%%%%%
with
$
 \Theta_1(y,\beta)=\cos^4\lmk\frac{\beta}2  \rmk \lmk j_0+\frac57
j_2+\frac{3}{112} j_4 \rmk ,
$
and 
$
 \Theta_2(y,\beta)=\lmk -\frac38 j_0+\frac{45}{56}
j_2-\frac{169}{896} j_4 \rmk
+\lmk \frac12 j_0-\frac57j_2-\frac{27}{224}j_4  \rmk \cos\beta \nonumber\\
+ \lmk-\frac18 j_0-\frac5{56}j_2-\frac{3}{896}j_4  \rmk \cos(2\beta).
$
The function $j_n$ is the $n$-th spherical Bessel function with its argument
$
y=f/f_D.% ={4\pi f R_E\sin(\beta/2)}/{c}.
$
%The expression (\ref{gi}) coincides with a corresponding formula  in 

\begin{figure}[!tb]
\begin{center}
\epsfxsize=6.cm
\begin{minipage}{\epsfxsize} \epsffile{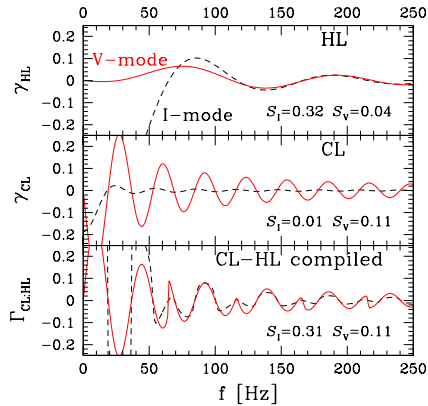} \end{minipage}
 \end{center}
\caption{Overlap  functions for the un-polarized $I$ mode (dashed
 curves), and the circularly polarized $V$-mode (solid curves). The
 upper panel shows 
 the results for the Hanford-Livingston (HL) pair (the characteristic
 frequency $f_D=100$Hz). The middle one
 is results for the LCGT-Livingston (CL) pair ($f_D=31$Hz). The
 normalized SNRs ${\it S}_{I,V}$ (with the adv LIGO noise spectrum) are 
also presented.  The bottom one show the compiled
 functions $\Gamma_{I,V}$ (eq.(\ref{co})) made from  both pairs.  
\label{fig:overlap} }
\end{figure}

\cite{Flanagan:1993ix}. 
On the other hand, the overlap function for the $V$-mode is
given by 
%%%%%%%%%%%%%%%%%%%%%%%%%%%%%%%%%%%%%%%%%%%%%%%%%%%%%%%%%%%%%%%%%%%%%%%%%%
\beq
\gamma_{V,ab}=\Theta_3(y,\beta)\,\sin(4\Delta) 
\label{gv}
\eeq
%%%%%%%%%%%%%%%%%%%%%%%%%%%%%%%%%%%%%%%%%%%%%%%%%%%%%%%%%%%%%%%%%%%%%%%%%%
with
$
\Theta_3(y,\beta)=-\sin\lmk \frac{\beta}2 \rmk \lkk \lmk-j_1+\frac78
j_3  \rmk + \lmk j_1+\frac38 j_3 \rmk\cos\beta   \rkk.
$
In Fig.~\ref{fig:overlap},  
the overlap functions for the two representative pairs are 
shown in top (HL) and middle (CL) panels.

Here, we give a simple interpretation for the angular 
dependence of Eqs.~(\ref{gi}) and (\ref{gv}).
The beam pattern functions $F^P_{a}$ and $F^P_{b}$ are given
by  linear combinations of $(\,\cos(2\sigma_1),~\sin(2\sigma_1)\,)$ and
$(\,\cos(2\sigma_2),~\sin(2\sigma_2)\,)$ respectively, reflecting their 
spin-2 like nature. Then, with Eq.~(\ref{gi1}) and addition formulas of
trigonometric functions, the overlap functions should be linear
combinations of $\cos[2(\sigma_1\pm \sigma_2)]$ and $\sin[2(\sigma_1\pm
\sigma_2)]$, namely, $\cos(4\Delta)$, $\cos(4\delta)$, $\sin(4\Delta)$ 
and $\sin(4\delta)$. Since the expectation value $C_{ab}(f)$ is a real
function for our beam 
pattern functions, we have $\lla s_a s_b^*\rra=\lla s_b s_a^*\rra$. This
essentially results in replacing the roles of $\sigma_1$ and
$\sigma_2$, and the functions $\gamma_I$ and $\gamma_V$ cannot contain
terms proportional to $\sin(4\delta)= \sin[2(\sigma_1-\sigma_2)]$.

On the other hand, while the observable $C_{ab}(f)$ and the amplitude $I$ 
are invariant under the parity transformation of a coordinate system, 
the sign of the parameter $V$ flips, because the transformation 
interchanges right-and left-handed waves. Therefore, the function 
$\gamma_{V,ab}$ must change its sign while keeping the quantity 
$C_{ab}(f)$ invariant. Geometrically,   
this corresponds to the re-definition of the azimuthal angles 
$\sigma_{1,2}$ in a clockwise direction (or putting $\sigma_1\to
-\sigma_1$ and $\sigma_2\to -\sigma_2$). As a result, the function
$\gamma_{V,ab}$ should be odd functions of $\delta$ and $\Delta$, and it
must be proportional to $\sin (4\Delta)$ as shown in Eq.~(\ref{gv}) (the term
proportional to $\sin(4\delta)$ is already prohibited as explained 
earlier). With similar arguments,  we find that  the 
function $\gamma_I$ is a linear combination of $\cos (4\Delta)$
and $\cos (4\delta)$ as in Eq.~(\ref{gi}).

%-%-%-%-%-%-%-%-%-%-%-%-%-%-%-%-%-%-%-%-%-%-%-%-%-%-%-%-%-%-%-%-%-%-%-%-%-
%-%-%-%-%-%-%-%-%-%-%-%-%-%-%-%-%-%-%-%-%-%-%-%-%-%-%-%-%-%-%-%-%-%-%-%-%-
\subsection{Special cases}
%\underline{\em A)  Special cases}
%-%-%-%-%-%-%-%-%-%-%-%-%-%-%-%-%-%-%-%-%-%-%-%-%-%-%-%-%-%-%-%-%-%-%-%-%-
%-%-%-%-%-%-%-%-%-%-%-%-%-%-%-%-%-%-%-%-%-%-%-%-%-%-%-%-%-%-%-%-%-%-%-%-%-

To stress the importance of the geometric configuration,  
it is instructive to consider several simple examples for idealistic pair of 
detectors. When a pair of detectors is co-located ($\beta=0^{\circ}$ 
and $D=0$), the functions $(\Theta_1,\Theta_2)$ defined after Eq.~(\ref{gi}) 
 become $(1,0)$ and we have $\gamma_{I,ab}=\cos(4\delta)$. 
The identity $\Theta_2=0$ at $\beta=0^{\circ}$ implies that 
the function $\gamma_I$ depends very weakly on the parameter 
$\Delta$ at a small angle $\beta \sim 0^{\circ}$. 
On the other hand, the overlap function $\gamma_{V,ab}$ always 
vanishes for a pair of detectors in the same plane $(\beta=0)$. 
This is even true with a finite separation $D\ne0$.  
This exact cancellation comes from 
the geometric symmetry of the beam pattern function with respect 
to the detector plane \cite{Seto:2006hf,Kudoh:2005as}.

For two detectors at antipodal positions ($\beta=180^\circ$), we have
$\Theta_1=0$ and the angle $\delta$ becomes geometrically 
meaningless. One can expect that the function 
$\gamma_{I,ab}$ is almost proportional to $\cos(4\Delta)$ near $\beta=180^\circ $.

%%%%%%%%%%%%%%%%%%%%%%%%%%%%%%%%%%%%%%%%%%%%%%%%%%%%%%%%%%%%%%%%%%%%%%%%%%
%%%%%%%%%%%%%%%%%%%%%%%%%%%%%%%%% Figure 2 %%%%%%%%%%%%%%%%%%%%%%%%%%%%%%%
%%%%%%%%%%%%%%%%%%%%%%%%%%%%%%%%%%%%%%%%%%%%%%%%%%%%%%%%%%%%%%%%%%%%%%%%%%
%%%%%%%%%%%%%%%%%%%%%%%%%%%%%%%%%%%%%%%%%%%%%%%%%%%%%%%%%%%%%%%%%%%%%%%%%%
%%%%%%%%%%%%%%%%%%%%%%%%%%%%%%%%%%%%%%%%%%%%%%%%%%%%%%%%%%%%%%%%%%%%%%%%%%

%-%-%-%-%-%-%-%-%-%-%-%-%-%-%-%-%-%-%-%-%-%-%-%-%-%-%-%-%-%-%-%-%-%-%-%-%-
%-%-%-%-%-%-%-%-%-%-%-%-%-%-%-%-%-%-%-%-%-%-%-%-%-%-%-%-%-%-%-%-%-%-%-%-%-
\subsection{Broadband SNR}
%\underline{\em B)  Broadband SNR}
%-%-%-%-%-%-%-%-%-%-%-%-%-%-%-%-%-%-%-%-%-%-%-%-%-%-%-%-%-%-%-%-%-%-%-%-%-
%-%-%-%-%-%-%-%-%-%-%-%-%-%-%-%-%-%-%-%-%-%-%-%-%-%-%-%-%-%-%-%-%-%-%-%-%-

Now, we turn to focus on a broadband sensitivity to the $I$- and $V$-modes. 
In the weak signal limit, the total signal-to-noise ratio (SNR) 
for the correlation signal $C_{ab}(f)$ is given by \cite{Flanagan:1993ix}
%%%%%%%%%%%%%%%%%%%%%%%%%%%%%%%%%%%%%%%%%%%%%%%%%%%%%%%%%%%%%%%%%%%%%%%%%%
\beq
{\rm SNR}^2=\lmk\frac{3H_0^2}{10\pi^2}  \rmk^2 T_{\rm obs} \lkk  
2\int_0^\infty df
\frac{X^2}{f^6 N_a(f)N_b(f)}  \rkk \label{broad}
\eeq
%%%%%%%%%%%%%%%%%%%%%%%%%%%%%%%%%%%%%%%%%%%%%%%%%%%%%%%%%%%%%%%%%%%%%%%%%%
with $X=\gamma_{I}\,\Omega_{\rm\scriptscriptstyle GW}+
\gamma_{V}\,\Omega_{\rm\scriptscriptstyle GW}\Pi$. We denote 
the noise spectra for detectors $a$ and $b$ by $N_a(f)$ and $N_b(f)$, 
assuming no noise correlation between them. 
In what follows, for simplicity of our analysis, 
we further assume that 
all the detectors have the same sensitivity comparable to 
the noise spectral curves of  advanced LIGO. The analytical fit 
from Fig.~1 of Ref.\cite{adv} leads to 
$
N(f)= 10^{-44} \lmk {f }/{\rm 10 Hz} \rmk^{-4}+10^{-47.25 }
\lmk {f }/{\rm 10^2 Hz} \rmk^{-1.7} {\rm Hz^{-1}}$ for $10\,{\rm Hz}\le
f \le 240\,{\rm Hz}$,  $N(f)=10^{-46} \lmk {f }/{\rm 10^3 Hz}
\rmk^{3}    ~~ {\rm Hz^{-1}}$ for  $240\,{\rm Hz}\le
f \le 3,000\,{\rm Hz}$, and otherwise $N(f)=\infty$. Note that 
the combination $f^6 N(f)^2$ becomes 
minimum around $f\sim 50$Hz with its bandwidth $\Delta f\sim 100$Hz. 
For a pair of coincident detectors 
(i.e., $\gamma_{I,ab}=1$ and $\gamma_{V,ab}=0$), 
the total SNR is evaluated by setting 
$X=\Omega_{\rm\scriptscriptstyle GW}$ in Eq.~(\ref{broad}), and we 
numerically obtain $
{\rm SNR}_0=4.8\,\lmk {T_{\rm obs}}/{3\,\rm yr}  \rmk^{1/2}\lmk
{\Omega_{\rm\scriptscriptstyle GW}h_{70}^2}/{10^{-9}}  \rmk.  
$

The total SNR depends strongly on model parameters of the background, 
including the polarization degree $\Pi$. 
In order to present our numerical results concisely, we first calculate 
${\rm SNR}_{\{I,V\},ab}$ by plugging $X=\gamma_{\{I,V\},ab}$ into 
Eq.~(\ref{broad}) and then normalize them as 
${\it S}_{\{I,V\},ab}\equiv {\rm SNR}_{\{I,V\},ab}/{\rm SNR}_0$.
The normalized SNRs can be regarded as  rms values of 
$\gamma_{\{I,V\},ab}$ with a weight function $[f^6\, N(f)^2]^{-1}$.

%-%-%-%-%-%-%-%-%-%-%-%-%-%-%-%-%-%-%-%-%-%-%-%-%-%-%-%-%-%-%-%-%-%-%-%-%-
%-%-%-%-%-%-%-%-%-%-%-%-%-%-%-%-%-%-%-%-%-%-%-%-%-%-%-%-%-%-%-%-%-%-%-%-%-
\subsection{Optimal configuration}
%\underline{\em C)  Optimal configuration}
%-%-%-%-%-%-%-%-%-%-%-%-%-%-%-%-%-%-%-%-%-%-%-%-%-%-%-%-%-%-%-%-%-%-%-%-%-
%-%-%-%-%-%-%-%-%-%-%-%-%-%-%-%-%-%-%-%-%-%-%-%-%-%-%-%-%-%-%-%-%-%-%-%-%-

Let us discuss optimal configurations of two detectors $(a,b)$ for
measuring the $I$- and $V$-modes with the correlation signal $C_{ab}$. 
There are two relevant issues here: maximization of the 
signals ${\it S}_{I,ab}$ and ${\it S}_{V, ab}$, 
and switching off either of them (${\it S}_{I,ab}=0$ or 
${\it S}_{I,ab}=0$) for their decomposition. To deal with the situation
comprehensively, we consider how to set the second detector  $b$ relative to
the fixed first one $a$ with a given separation angle $\beta$. 
In this case, the sensitivity to the $I$- and $V$-modes 
is characterized by the remaining adjustable parameters, 
$\sigma_1$ and $\sigma_2$. 
The former determines the position of the detector $b$, while the latter 
specifies its orientation (see Fig.~\ref{fig:detector_config}). 
Based on the expressions (\ref{gi}) and (\ref{gv}), 
one finds that there are three possibilities for the optimal detector 
orientation:  
$\cos (4\Delta)=-\cos (4\delta)=\pm 1$ (type I) or $\cos (4\Delta)=\cos 
(4\delta)=\pm 1$ (type II) to maximize the normalized 
SNR ${\it S}_{I,ab}$ \cite{Flanagan:1993ix}, and   
$\cos{(4\Delta)}=\cos{(4\delta)}=0$ (type III) 
to erase the contribution from $I$-mode. For type I, the solutions 
of the two angles $\sigma_{1,2}$ are $\sigma_1=\sigma_2=45^\circ$ 
(mod $90^\circ$) and the detector $b$ must be sited in one 
of the two great circles passing through the detector $a$, 
parallel to one of the two arms. 
For type II, the second detector must reside in two great 
circles parallel or perpendicular to the bisecting line of 
each detector. Similarly, the type III configuration is 
realized by placing the second detector on one of the 
four great circles defined for types I and II, with 
rotating $45^{\circ}$ relative to the first detector.

Note that the sensitivity to the $V$-mode is automatically switched off 
for the type I and II configurations and is conversely maximized for 
the type III configuration. This is because the normalized SNR 
${\it S}_{V,ab}$ is proportional to $\sin (4\Delta)$. 
While a definite detection of a weak $V$-mode signal requires  
a careful removal of the $I$-mode signal from observed 
data, it turns out that the geometrical requirement for type III 
configuration is severe. 
As we see later, however, we can easily control the contribution 
from the $I$- (or $V$-)mode by introducing a third detector.

\begin{figure}[!bth]
\begin{center}
\epsfxsize=6.cm
\begin{minipage}{\epsfxsize} \epsffile{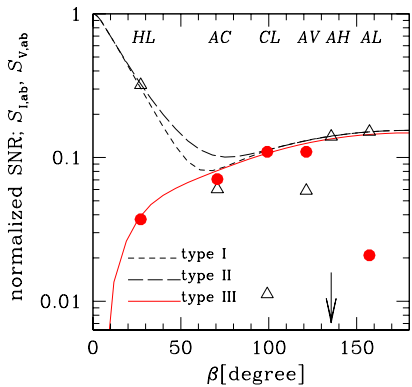} \end{minipage}
 \end{center}
\caption{ Normalized signal to noise ratios (${\it S}_{I,ab}$ and 
${\it S}_{V,ab}$) with optimal configurations  
for the $I$-mode (short dashed curve: type I, long dashed curve: type II)
 and for  the $V$-mode (solid curve: type III with setting $\Pi=1$ for
 illustrative purpose). We use the noise curve for
 the advanced LIGO.   For each detector pair,
 ${\it S}_I$ and ${\it S}_V$ are given with a  triangle and a  circle
 respectively at its separation $\beta$. There are four other pairs not
 shown here; CH
 with $(S_I, S_V)=(0.04,0.08)$, LV with (0.08,0.04), HV with (0.07,0.06)
 and CV with (0.09,0.04).
\label{fig:SNR}} 
\end{figure}

In Fig.~\ref{fig:SNR}, we present the normalized SNRs 
for the optimal geometries; types I, II and III (short-dashed,
long-dashed, and solid curves, respectively). One noticeable point is
that  
a widely separated  ($\beta\sim 180^\circ$) pair is powerful to search 
for the $V$-mode (recall the cancellation $\gamma_V=0$ at $\beta=0$). To
reduce the contribution from the $I$-mode,  
pairs that are usually disadvantageous to measuring the total intensity 
$\Omega_{\rm\scriptscriptstyle GW}$ now play a very  
important role. In Fig.~\ref{fig:SNR}, we also show the 
normalized SNRs for representative pairs made from the five detectors,  
in which several interesting combinations are found. 
The HL (with $\cos(4\delta)\sim 1$ and $\sin(4\Delta)\sim 0.93$)
realizes nearly maximum values simultaneously for ${\it S}_{I,ab}$ and 
${\it S}_{V,ab}$ at its separation $\beta=27.2^\circ$. 
This is because ${\it S}_{I,ab}$ is mainly determined by the angle 
$\delta$ at a small  $\beta$, while ${\it S}_{V,ab}$ depends only on 
$\Delta$. 
%We have $\cos 4\Delta\sim 1$ and $\sin 4\Delta\sim 0.93$ for HL (see
%Table 2). 
The CL has good sensitivity to the $V$-mode and relatively insensitive to 
the $I$-mode with $\sin(4\Delta)\sim1$. 
In contrast, AH is almost insensitive to the $V$ mode with 
$\sin 4\Delta=-0.007$. In this sense, 
LCGT and AIGO detectors are suitably oriented
to probe the $I$- and $V$-modes, respectively.

%%%%%%%%%%%%%%%%%%%%%%%%%%%%%%%%%%%%%%%%%%%%%%%%%%%%%%%%%%%%%%%%%%%%%%%%%%
%%%%%%%%%%%%%%%%%%%%%%%%%%%%%%%%% Figure 3 %%%%%%%%%%%%%%%%%%%%%%%%%%%%%%%
%%%%%%%%%%%%%%%%%%%%%%%%%%%%%%%%%%%%%%%%%%%%%%%%%%%%%%%%%%%%%%%%%%%%%%%%%%
%%%%%%%%%%%%%%%%%%%%%%%%%%%%%%%%%%%%%%%%%%%%%%%%%%%%%%%%%%%%%%%%%%%%%%%%%%
%%%%%%%%%%%%%%%%%%%%%%%%%%%%%%%%%%%%%%%%%%%%%%%%%%%%%%%%%%%%%%%%%%%%%%%%%%

%-%-%-%-%-%-%-%-%-%-%-%-%-%-%-%-%-%-%-%-%-%-%-%-%-%-%-%-%-%-%-%-%-%-%-%-%-
%-%-%-%-%-%-%-%-%-%-%-%-%-%-%-%-%-%-%-%-%-%-%-%-%-%-%-%-%-%-%-%-%-%-%-%-%-
\subsection{Separating $I$- and $V$-modes}
%\underline{\em D)  Separating $I$- and $V$-modes}
%-%-%-%-%-%-%-%-%-%-%-%-%-%-%-%-%-%-%-%-%-%-%-%-%-%-%-%-%-%-%-%-%-%-%-%-%-
%-%-%-%-%-%-%-%-%-%-%-%-%-%-%-%-%-%-%-%-%-%-%-%-%-%-%-%-%-%-%-%-%-%-%-%-%-

As a final mention, we will address the issue of $I$- and $V$-mode separation 
by combining several pairs of detectors. 
For preliminary investigation, we consider the case that two pairs of 
interferometers 
$(a,b)$ and $(c,d)$ are available. Detectors $a$ and $c$ can be identical, 
but we need at least three independent detectors for the study below. 
First note that the correlation signals are given by 
$C_{ab}(f)=\gamma_{I,ab}(f)\,I(f)+\gamma_{V,ab}(f)\,V(f)$ and 
$C_{cd}(f)=\gamma_{I,cd}(f)\,I(f)+\gamma_{V,cd}(f)\,V(f)$. 
From this, one can easily find that the contribution from the 
$I$-mode is canceled by taking a 
combination $W\equiv\gamma_{I,ab}\,C_{cd}-\gamma_{I,cd}\,C_{ab}=
(\gamma_{V,cd}\,\gamma_{I,ab}-\gamma_{V,ab}\,\gamma_{I,cd})\,V(f)$. 
The statistical analysis based on the combination $W$ would be a robust 
approach for actual $V$-mode search, although a further 
refinement may be possible by combining more pairs, 
which we will report elsewhere. 

Since the rms amplitude of the detector noise for the combination $W$
becomes $N(f)\,(\gamma_{I,ab}^2+\gamma_{I,cd}^2)^{1/2}$, we define the 
{\it compiled} overlap function for the $V$ mode by 
%%%%%%%%%%%%%%%%%%%%%%%%%%%%%%%%%%%%%%%%%%%%%%%%%%%%%%%%%%%%%%%%%%%%%%%%%%
\beq
\Gamma_{V,ab:cd}\equiv
\frac{\gamma_{V,cd}\gamma_{I,ab}-\gamma_{V,ab}\gamma_{I,cd}}{[\gamma_{I,ab}^2 
+\gamma_{I,cd}^2]^{1/2}}.\label{co}
\eeq
%%%%%%%%%%%%%%%%%%%%%%%%%%%%%%%%%%%%%%%%%%%%%%%%%%%%%%%%%%%%%%%%%%%%%%%%%%
This expression should be used  in Eq.~(\ref{broad}) when evaluating 
the broadband SNR for the $V$-mode with the combination $W$. 
%This function is proportional to the area of a triangle shaped by two
%vectors 
%$(\gamma_{I,ab},\gamma_{I,cd})/[\gamma_{I,ab}^2+\gamma_{I,cd}^2]^{1/2}
%$ and  $(\gamma_{V,ab},\gamma_{V,cd})$ with  $|\Gamma_{V,ab:cd}|\le 
%[\gamma_{V,ab}^2 %+\gamma_{V,cd}^2]^{1/2}$. 
%This sounds reasonable since the
%right-hand-side represents the simple summation of two independent
%sensitivities. 
In a similar way, we define the compiled function 
$\Gamma_{I,ab:cd}$ for the $I$ mode by 
interchanging the subscripts $V$ and $I$ in Eq.~(\ref{co}). 
Bottom panel of Fig.~\ref{fig:overlap} shows the compiled overlap functions 
$\Gamma_{\{I,V\},ab:cd}$ from two pairs of detectors, CL-HL.  
% ???Based on the
%geometric interpretation for the compiled functions, we intend to use CL
%to mainly extract the $V$ mode and HL for the $I$ mode.??? 
With this combination, the normalized SNR becomes $0.11$ for the $V$-mode 
and $0.31$ for the $I$-mode. 
Using numerical results below eq.(\ref{broad}), the detection limit for
the 
polarization degree $\Pi$  is given as
$\Pi= (T/3{\rm yr})^{-1/2}(SNR_V/5)(\Omega_{\rm\scriptscriptstyle
GW}h_{70}^2/10^{-8})^{-1}$ with signal-to-noise ratio $SNR_V$. 
These numerical results are almost the same 
values as in ${\it S}_V$ for CL and ${\it S}_I$ for HL, and in this
sense,  
the $I$-, $V$-mode separation can be performed efficiently with 
naively expected sensitivities ${\it S}_{\{I,V\},ab}$. 
%Note that the sensitivity to 
%the $V$-mode by the HL-CL network is about three times worse than 
%the $I$-mode probed by HL. 
Note  that the other combinations, such as AV-HL, AV-HV and CL-HV, 
also provide the normalized value $\sim0.11$ for the $V$-mode, 
but AH-AL has only $0.015$.

In summary, we  reported principle aspects for measuring a
circular polarization degree of a gravitational wave background that is
related to parity violation. We find that pairs of ground-based 
interferometers that are widely separated and relatively insensitive to the 
total intensity $\Omega_{\rm\scriptscriptstyle GW}$ are advantageous 
for the measurement. With at 
least three detectors, the polarization degree and the intensity 
$\Omega_{\rm\scriptscriptstyle GW}$ can be separately detected.

We would like to thank N. Kanda and M. Ando for supplying information on LCGT 
and comments. This work was supported in part by  a Grant-in-Aid for Scientific Research from the Japan Society for the 
Promotion of Science (No.~18740132).

\end{document}